# All-Optically Controlled Memristive Reservoir Computing Capable of Bipolar and Parallel Coding

Lingxiang Hu[1,#], Dian Jiao[1,#], Kexuan Wang[1], Peihong Cheng[1,2], Jingrui Wang[1,2], Li Zhang[3], Athanasios V. Vasilakos[4], Yang Chai[5], Zhizhen Ye[6,7] & Fei Zhuge[1,6,8,9] *

[1]Ningbo Institute of Materials Technology and Engineering, Chinese Academy of Sciences, Ningbo, China

[2]School of Electronic and Information Engineering, Ningbo University of Technology, Ningbo, China

[3]Healthcare Engineering Centre, School of Engineering, Temasek Polytechnic, Singapore

[4]Center for AI Research, University of Agder, Grimstad, Norway

[5]Department of Applied Physics, The Hong Kong Polytechnic University, Hong Kong, China

[6]Institute of Wenzhou, Zhejiang University, Wenzhou, China

[7]State Key Laboratory of Silicon and Advanced Semiconductor Materials, School of Materials Science and Engineering, Zhejiang University, Hangzhou, China

[8]Center for Excellence in Brain Science and Intelligence Technology, Chinese Academy of Sciences, Shanghai, China

[9]College of Materials Science and Opto-Electronic Technology, University of Chinese Academy of Sciences, Beijing, China

[#]L.H. and D.J. contributed equally to this paper.
*E-mail: zhugefei@nimte.ac.cn

**Abstract:**

　　Physical reservoir computing (RC) utilizes the intrinsic dynamical evolution of physical systems for efficient data processing. Emerging optoelectronic RC platforms – such as light-driven memristors – merge the benefits of electronic and photonic computation. However, conventional designs are often limited by the unipolar photoresponse of optoelectronic devices, which restricts reservoir state diversity and reduces computational accuracy. To overcome these limitations, we introduce an all-optically controlled RC system employing an oxide memristor array that demonstrates exceptional uniformity and stability. The memristive devices exhibit wavelength-dependent bipolar photoresponse, originating from light-induced dynamic evolution of oxygen vacancies. Tuning the power density and irradiation mode of dual-wavelength light pulses enables dynamic control of photocurrent relaxation and nonlinearity. By leveraging these unique device properties, we develop bipolar and parallel coding strategies to significantly enrich reservoir dynamics and enhance nonlinear mapping capability. In word recognition and time-series prediction tasks, the bipolar coding demonstrates markedly improved accuracy compared to unipolar coding. The parallel coding supports multi-source signal fusion within a single reservoir, maintaining high computational accuracy while significantly reducing hardware consumption. This work provides a high-performance approach to physical RC, paving the way for intelligent edge computing.

**Introduction**

With the rapid advancement of the Internet of Things and intelligent terminal devices, edge computing – a distributed computing paradigm – is garnering growing attention[1]. Its primary goal is to offload computational tasks to front-end hardware at the point of data generation, thereby reducing latency, conserving bandwidth, and improving real-time system performance. In this context, there is an increasing demand to deploy lightweight artificial neural networks (ANNs) on front-end hardware for efficient sensor data processing[2]. Reservoir computing (RC), a computational framework inspired by recurrent neural networks, replaces the conventional trainable recurrent layer with a fixed, non-linear reservoir[3]. This reservoir acts as a dynamical system, projecting input signals into a high-dimensional computational space, making RC highly efficient for processing temporal signals. As only the output layer weights require training, the framework significantly reduces computational complexity and hardware resource requirements, rendering it a highly promising technology for edge intelligence[3,4].

To further enhance the energy efficiency of RC systems for resource-efficient information processing, researchers have increasingly turned to physical hardware implementations – leveraging the inherent dynamical evolution of physical systems to achieve efficient, high-dimensional mapping of input signals[5,6]. Various physical platforms, including electronics[6–10], optical[11–13], and mechanical devices[14,15], have been explored for implementing RC in hardware. Among these, emerging memory devices like memristors[7,9,16–19], which exhibit intrinsic nonlinear dynamics and input-dependent memory effects, can naturally emulate neuronal nodes within the reservoir, enabling efficient signal processing directly at the hardware level. Such memristive RC implementations demonstrate superior energy efficiency compared to software-based RC on conventional digital computing systems, providing a promising pathway for deployment in edge environments[16,18].

Optoelectronic RC, exemplified by novel light-driven memristors[20–24], combines the advantages of both electronics and photonics. On one hand, the inputs in optoelectronic RC systems are optical signals, benefiting from high bandwidth, low crosstalk, and parallel processing capabilities. On the other hand, the outputs are electrical signals, ensuring full compatibility with conventional CMOS integrated circuits. Furthermore, optoelectronic RC can be seamlessly integrated with in-sensor computing technology[25–29], merging perception and computation into a unified process. This enables preliminary signal processing directly at the sensor level, reducing data transmission requirements and thereby lowering system power consumption and latency. Optoelectronic RC has achieved remarkable progress in recent years, especially in tasks such as multimodal recognition[30,31] and chaotic system prediction[32,33].

The performance of RC systems hinges on the mapping capability of their nonlinear nodes, where richer nodal dynamics directly boost computational performance[6,7,9]. Conventional optoelectronic RC devices, however, are typically limited by unipolar photoresponses[20,21,23–33]. These monotonic saturating responses constrain dynamic range and nonlinearity of the reservoir. Consequently, the diversity of internal reservoir states is reduced, and strong input signals can lead to device failure, ultimately degrading performance on complex temporal tasks[6]. To overcome these challenges, researchers have introduced additional dynamic electrical control signals and adopted hybrid optoelectronic coding strategies that markedly enhance RC accuracy[30,31]. Nevertheless, these approaches also increase system complexity and impose additional energy costs.

Beyond single-signal processing, advanced edge intelligence also requires the ability to handle multi-source information efficiently[34]. By integrating signals with multiple dimensions and features from different sensors, multi-source information fusion mitigates errors associated with any single source[35]. However, existing physical RC systems are typically specialized for single-source signal processing through a serial coding mode[5,6]. As a result, handling multi-source tasks often requires multiple independent reservoirs operating in parallel[18,28,36]. On the contrary, parallel coding facilitates efficient multi-source information processing by allowing synchronous mapping of diverse signals into a single reservoir, thereby combining feature extraction with information fusion and minimizing resource consumption[35,37]. Nevertheless, implementing parallel coding of multi-source signals within the same reservoir device remains challenging, requiring a tightly coupled physical mechanism that enables direct interaction between different signals. Thus, a critical need remains for a single, simple device platform that intrinsically provides rich nonlinear dynamics for high-accuracy RC and natively supports the efficient processing of multi-source signals, without external control complexity.

To address these challenges, this work introduces the concept of all-optically controlled reservoir computing (AOC RC) – where both the programming and the fundamental neuromorphic dynamics are governed solely by optical signals, eliminating the need for external electrical control circuits. We propose that all-optically controlled memristors (AOCMs)[38–43] and analogous bipolar optoelectronic devices[44–48] can be effectively employed to construct AOC RC hardware systems. These devices exhibit a bidirectional photoresponse, allowing reversible modulation of conductance states by adjusting the wavelength or power density of incident light, thereby facilitating the emulation of more complex neural dynamics. This capability substantially expands the reservoir's effective state space, enhances its nonlinear mapping ability, and ultimately improves RC accuracy and robustness. Furthermore, by leveraging the bipolar optoelectronic properties of these devices, we propose novel bipolar and parallel coding strategies based on optical signals for efficient information processing within a physical reservoir.

Implementing this concept, we fabricate a highly stable 16×16 array of ZnO-based AOCMs. After ultraviolet (UV) light pretreatment, the devices exhibit positive persistent photoconductivity (PPPC) under blue light illumination and negative persistent photoconductivity (NPPC) under red light illumination – a response originating from the dynamic evolution of neutral and ionized oxygen vacancies under different wavelengths of light. Furthermore, under alternating or simultaneous exposure to two wavelengths, the devices demonstrate dynamically tunable bipolar photoresponse behaviors. This intrinsic property enables us to demonstrate, for the first time in a single device platform, both bipolar and parallel coding schemes. When applied to English word recognition and time-series prediction, our AOC RC system with bipolar coding achieves 93% accuracy and a normalized root mean square error (NRMSE) of 0.12, respectively, – a dramatic improvement over unipolar coding (76% accuracy and an NRMSE of 0.31). Most notably, for multi-dimensional recognition, our parallel coding scheme integrates feature extraction and multi-signal fusion within the same reservoir, maintaining high accuracy (90.5%) comparable to that of bipolar coding, while substantially reducing hardware resource overhead.

**Physical RC based on AOCMs**

**Figure 1a** illustrates the basic structure of an RC network: an input layer, a reservoir layer exhibiting complex time-dependent dynamics, and a linear readout layer. The reservoir layer's role is to nonlinearly project low-dimensional inputs into a high-dimensional feature space, represented by the neuron states. Physical RC seeks to exploit physical dynamics as the computing resources for nonlinear mapping within the RC framework, enabling resource-efficient information processing.

The diversity and richness of the reservoir states are essential for the precision of this nonlinear mapping. This richness is fundamentally determined by the dynamic range and linearity of the physical nodes themselves. As **Fig. 1b** illustrates, traditional optoelectronic RC, reliant on unidirectional photoresponse devices, offers a limited state space. Our proposed AOC RC, in contrast, not only merges the advantages of electronic and photonic RC but – crucially – leverages bidirectional photoresponses to provide a significantly richer and more expressive reservoir state space.

To achieve a stable and uniform hardware platform, we fabricated a Pt/ZnO/Pt memristor array (see **Methods** for details). During device fabrication, a key post-annealing treatment was applied to the ZnO thin film to relieve internal stress and enhance crystallinity, which significantly improved device-to-device uniformity and operational stability – critical factors for reliable array operation. Detailed optoelectronic characterization of the array will be presented in the following section. **Figure 1c** shows a top-view optical microscope image of a typical 16×16 crossbar array, along with magnified views of its 3×3 sub-arrays. **Supplementary Fig. 1** illustrates the current–voltage characteristics of the device under dark conditions, revealing pronounced self-rectifying behavior. This characteristic, likely originating from asymmetric Schottky barriers formed between the ZnO layer and the two electrodes, effectively suppresses sneak current paths and minimizes crosstalk within the crossbar array. After the post-annealing treatment, a strong Schottky junction is formed between ZnO and the bottom Pt electrode. Meanwhile, due to oxygen or moisture adsorption from ambient air on the ZnO surface, the Schottky barrier height at the ZnO/top Pt electrode interface is significantly reduced[49]. These two asymmetric Schottky junctions lead to the observed self-rectifying behavior.

The core enabling functionality of our system is shown in **Fig. 1d**. Our ZnO-based AOCMs exhibit a wavelength-dependent bipolar photoresponse (*i.e.*, PPPC and NPPC), offering rich reservoir states for RC. A constant read voltage ($V_r = 1$ mV) was used to monitor real-time changes in device current ($\Delta I$). Compared to conventional physical RC with unipolar and serial coding, our AOCM-based RC hardware achieves highly efficient data processing through unique bipolar and parallel coding (**Fig. 1e**).

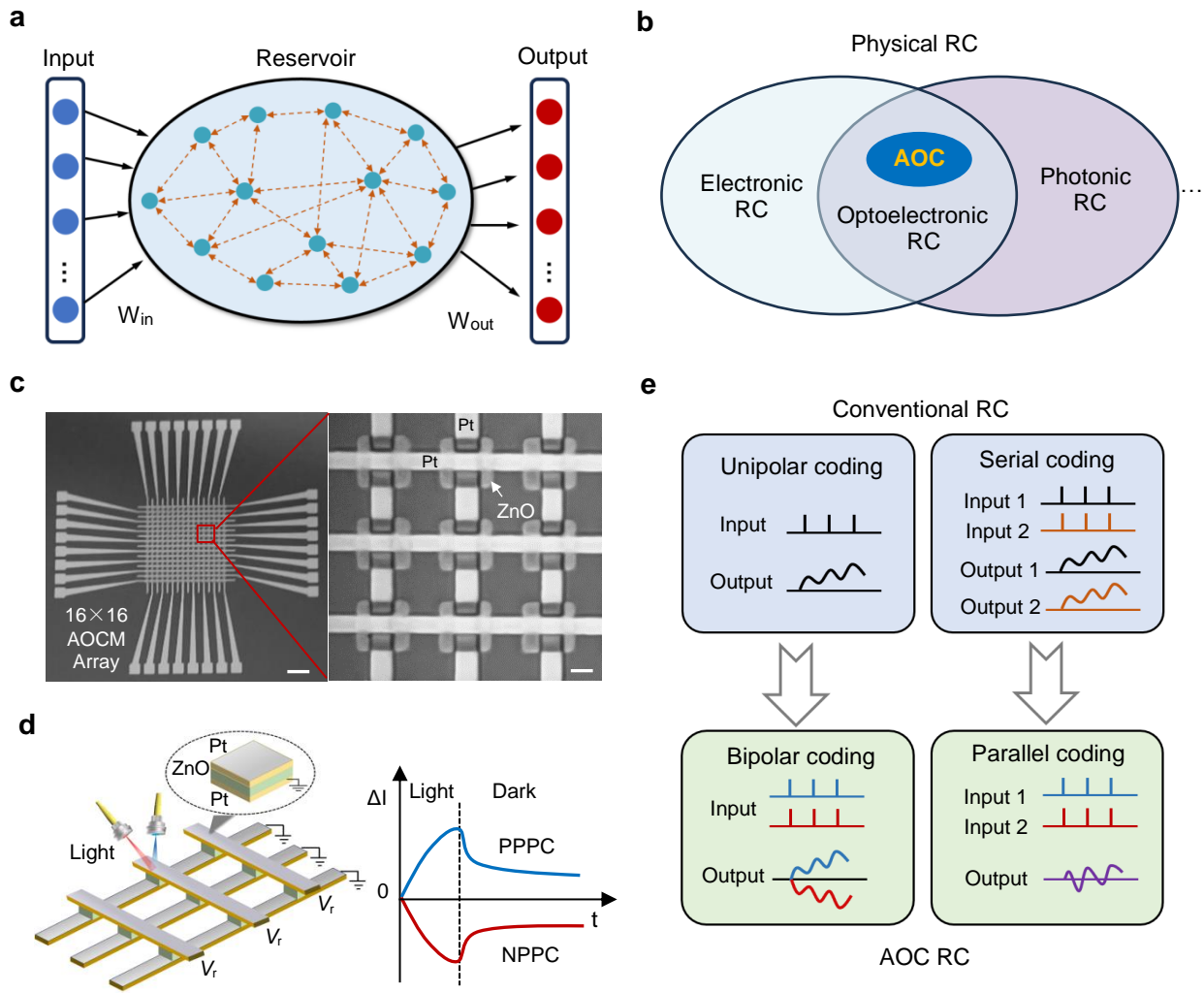

**Fig. 1 | AOC RC based on AOCM array. a**, Schematic of a RC system. The input is fed to a reservoir composed of a large number of nonlinear nodes. The recognition tasks of input information are performed by training the output weights. **b**, General classification of physical RC, illustrating the relationships among electronic RC, photonic RC, optoelectronic RC, and AOC RC. **c**, Optical microscope image of a 16×16 ZnO AOCM crossbar array (scale bar: 1 mm), along with magnified views of its 3×3 sub-array (scale bar: 50 μm). **d**, Schematic diagram of the AOCM array operation mode. Different wavelengths of light are separately irradiated onto the device. A constant read voltage ($V_r$ = 1 mV) is used for real-time monitoring of current changes (ΔI). ΔI is defined as the difference between current values measured before and after illumination. The AOCMs show a bipolar photoresponse (*i.e.*, PPPC and NPPC). **e**, Comparison of AOC RC and conventional RC in coding mode.

**Bipolar Photoresponse of AOCMs**

Wavelength-programmable bipolar photoresponse is the key characteristic enabling AOC RC. This capability to modulate conductance all-optically in both positive and negative directions provides the essential nonlinear dynamics and expanded state space that distinguish AOC RC from conventional unipolar systems. Initially, the ZnO memristor shows a pronounced positive response under UV light illumination (**Fig. 2a**). Interestingly, after exposure to UV light, the device exhibits bipolar photoresponses to blue (*e.g.*, 405 nm) and red (*e.g.*, 650 nm) light, demonstrating positive and negative responses, respectively, as illustrated in the inset of **Fig. 2a**. As clarified in our previous work[40], the bipolar photoresponse mechanism of ZnO-AOCMs mainly originates from the competition between the ionization and neutralization of oxygen vacancies in ZnO under light illumination. Under short-wavelength light (*e.g.*, 405 nm), the ionization of oxygen vacancies dominates, increasing the device conductance and generating a positive photoresponse. In contrast, under long-wavelength light (*e.g.*, 650 nm), the neutralization of oxygen vacancies prevails, reducing device conductance and leading to a negative photoresponse.

**Figure 2b** depicts the dependence of the device's bipolar photoresponse on the power density ($P$) under 405 nm and 650 nm light illumination. As expected, the bipolar photoresponse progressively increases with rising power densities at both wavelengths. When illumination ceases, the photocurrent exhibits pronounced spontaneous relaxation – a key feature of optoelectronic devices employed in RC systems, as it inherently links current inputs to prior states. Conventional unipolar optoelectronic RC systems often exhibit limited tunability in temporal dynamics, mainly due to challenges in modulating the photocurrent relaxation time, which consequently restricts their information processing capability. The distinctive bipolar photoresponse of AOCM devices offers an effective means to overcome this limitation. By applying a secondary light source during the relaxation phase, the photocurrent decay dynamics can be actively modulated via controlled tuning of the oxygen vacancy ionization or neutralization rates. As illustrated in **Fig. 2c**, following the positive photoresponse induced by 405 nm light, subsequent exposure to 650 nm light accelerates the photocurrent decay relative to its natural relaxation, thereby enabling the modulation of relaxation speed by adjusting the 650 nm light power density. Similarly, following the negative photoresponse induced by 650 nm light, subsequent irradiation with 405 nm light can accelerate the photocurrent increase compared to its natural relaxation (**Fig. 2d**). **Figure 2e** displays the photocurrent distribution under various combinations of 405 nm and 650 nm light power densities, clearly illustrating that by adjusting the power densities of these two light sources, dynamic modulation of the relaxation behaviors in the bipolar photoresponse can be achieved. The light application schemes used here are the same as those in **Fig. 2c**. **Figure 2f** shows the distinct different photoresponses of the device when alternating 405 nm and 650 nm light pulses with different power density combinations are applied. These findings confirm that the bipolar photoresponse of our AOCMs exhibits rich relaxation dynamics suitable for bipolar coding RC.

As we know, light signals exhibit high parallelizability, enabling modulation of the device state by multi-beam light illumination. We observe that exposing the device to synchronized 405 nm and 650 nm light also produces diverse photoresponses and relaxation dynamics, as illustrated in **Fig. 2g–i**. It deserves noting that in **Fig. 2g**, compared to single-beam light illumination, dual-beam irradiation results in strongly nonlinear photoresponses with fluctuating up-and-down features, likely arising from the dynamic competition between oxygen vacancy ionization and neutralization under 405 nm and 650 nm illumination, respectively. The capability of our AOCMs to process parallel light signals lays the groundwork for implementing unique parallel coding RC.

Furthermore, the uniformity and stability of the 16×16 AOCM array were evaluated. **Figure 2j,k** displays the initial current maps and the corresponding statistical distribution for all 256 devices in the array, respectively. **Figure 2l,m** presents the ΔI maps and the corresponding statistical distribution after illumination with light of different wavelengths, respectively. These results demonstrate excellent device-to-device uniformity within the AOCM array. Moreover, the devices maintain remarkable performance stability over time. The bipolar photoresponse characteristics remain almost unchanged even after 9 months regardless of the light application modes (**Supplementary Fig. 2**). The exceptional uniformity and long-term operational stability of the devices are essential for achieving reliable bipolar and parallel coding RC.

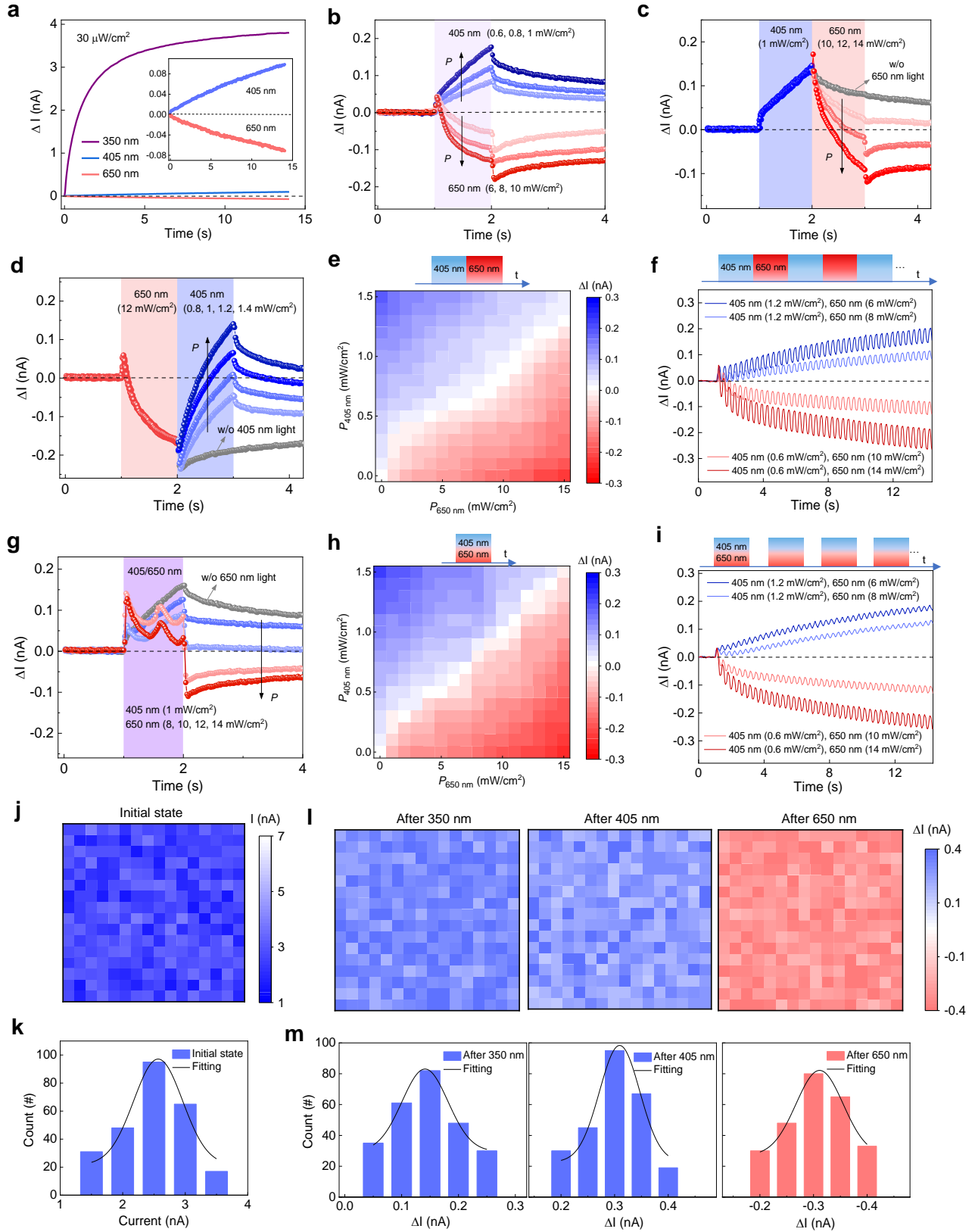

**Fig. 2 | Bipolar photoresponse of AOCMs. a,** Photoresponse upon irradiation with light of various wavelengths ($P$ = 30 μW cm$^{-2}$). The insets illustrate the enlarged views of the device's photoresponse under 405 nm and 650 nm light illumination. **b,** Bipolar photoresponse under

various power densities using independent illumination mode. **c,d** Relaxation behaviors of photocurrent under alternating irradiation mode. After exposure to a 405 nm light pulse with a fixed power density ($P$ = 1 mW/cm$^2$), the device is then irradiated with a 650 nm light pulse of gradually increasing power densities (**c**). The gray curve in the figure represents the photoresponse under 405 light irradiation only. Conversely, after a 650 nm light pulse with a fixed power density ($P$ = 12 mW/cm$^2$), the device is then exposed to a 405 nm light pulse with gradually increasing power densities (**d**). The gray curve represents the photoresponse under 650 light irradiation only. **e**, ΔI distribution for various combinations of $P$ under the illumination mode in (**c**). The statistical values of ΔI represent the photocurrents at the end of illumination. **f**, Bipolar photoresponse under alternating illumination with 405 nm (width ($W$) = 0.15 s, number ($N$) = 20) and 650 nm ($W$ = 0.15 s, $N$ = 20) light pulses by adjusting their power densities. In this mode, the two light pulses are alternately applied without any time interval. **g**, Photoresponse under synchronous irradiation mode. A 405 nm light pulse with a fixed power density ($P$ = 1 mW/cm$^2$) and a 650 light pulse with varying power densities are synchronously irradiated onto the device. The gray curve represents the photoresponse under 405 light irradiation only. Note that the $W$ of the 405 nm and 650 nm light pulses in (**b-d, g**) are all 1 s. **h**, ΔI distribution across different combinations of $P$ under synchronous illumination, similar to (**g**). The statistical values of ΔI represent photocurrents measured 0.3 s after the illumination ends. **i**, Bipolar photoresponse under synchronous illumination using 405 nm and 650 nm light ($W$ = 0.15 s, interval ($In$) = 0.15 s, $N$ = 40) by adjusting their power densities. **j**, Distribution maps of the initial current for all 256 devices in the 16×16 array. **k**, Statistical histograms of the initial current values in (**j**). **l**, Distribution maps of the ΔI under different light illuminations at 350 nm ($P$ = 30 μW/cm$^2$, $W$ = 0.2 s), 405 nm ($P$ = 1.2 mW/cm$^2$, $W$ = 1 s), and 650 nm ($P$ = 12 mW/cm$^2$, $W$ = 1 s). **m,** Statistical histograms of the corresponding ΔI values in (**k**). The results in (**k, m**) are well-fitted by a normal distribution curve. Note: In all tests of this work, pre-irradiation using 350 nm light ($P$ = 30 μW/cm$^2$, $W$ = 0.2 s) is needed before exposure to 405 or 650 nm light.

**Bipolar Coding for Word Recognition and Time-Series Prediction**

To demonstrate the bipolar coding capability of the proposed AOC RC, we conducted English word recognition tasks based on a 3×3 AOCM array. The task involved recognizing four-letter words, with each letter represented by a 3×3 pixel pattern. A total of 13 letters were used: 'C', 'H', 'I', 'J', 'K', 'X', 'Y', 'L', 'T', 'O', 'U', 'V', and 'Z', as depicted in **Supplementary Fig. 3**. These letters formed 20 distinct four-letter English words: 'COZY', 'HOLL', 'JOLT', 'KILO', 'LICK', 'LUCK', 'COOK', 'CITY', 'HOLY', 'OILY', 'YOLY', 'ITCH', 'TUCK', 'LOCK', 'COOL', 'TICK', 'HULK', 'KILL', 'VOLT', and 'TOXI' to rigorously test the discriminative power of different coding schemes. **Figure 3a** illustrates the operation mode of bipolar coding for word recognition, using 'CITY' as an example, where a bit '1' is mapped to a 405 nm (blue) light pulse, while a bit '0' to a 650 nm (red) light pulse. Critically, this scheme utilizes both positive and negative conductance modulation regimes of the AOCM, thereby doubling the effective state space of the reservoir compared to unipolar coding (**Fig. 3b**). For each word, nine parallel data streams were input into a 3×3 AOCM array reservoir and the resulting photocurrent states were fed into the readout layer for recognition (See **Methods**).

In bipolar coding schemes, selecting the appropriate power density combinations for the two light sources is crucial for achieving optimal coding performance. To fully utilize the bipolar coding space and enhance the distinguishability of reservoir states, these states should be distributed relatively uniformly across both positive and negative coding regions. As illustrated in **Fig. 2e**, suitable power density combinations lie along the diagonal region of the parameter space. In addition, the power densities should neither be too low nor too high. Excessively low power densities yield a weak photoresponse, limiting the distribution space of reservoir states and reducing their diversity. Conversely, although high power densities can generate a broader distribution of reservoir states, they tend to cause rapid photocurrent saturation, which negatively affects the correlation between sequentially input signals. Therefore, we selected power densities of 0.8 mW/cm² for 405 nm light and 10 mW/cm² for 650 nm light. This combination effectively balances the trade-off between reservoir state diversity and temporal signal correlation.

**Figure 3c** and **Supplementary Fig. 4** illustrate the evolution of photocurrent for each device in a 3×3 array in response to 16 distinct input combinations under bipolar coding. The resulting reservoir states are evenly distributed across both positive and negative coding regions, influenced by the optical stimulation history before and after each pulse. In contrast, unipolar coding (**Fig. 3d** and **Supplementary Fig. 5**) confines the reservoir states to the positive region, thereby reducing their separability. As shown in **Fig. 3e**, the recognition accuracy of four-letter words under bipolar coding reaches approximately 93% after 100 training epochs, whereas unipolar coding achieves only about 76%. The confusion matrices in **Figure 3f,g** highlight this advantage for highly similar words. For instance, the confusion probability between 'LOCK' and 'LUCK' is 48% under unipolar coding but drops to just 5% under bipolar coding. This confirms that the expanded, signed state space of bipolar coding directly translates to superior feature discrimination.

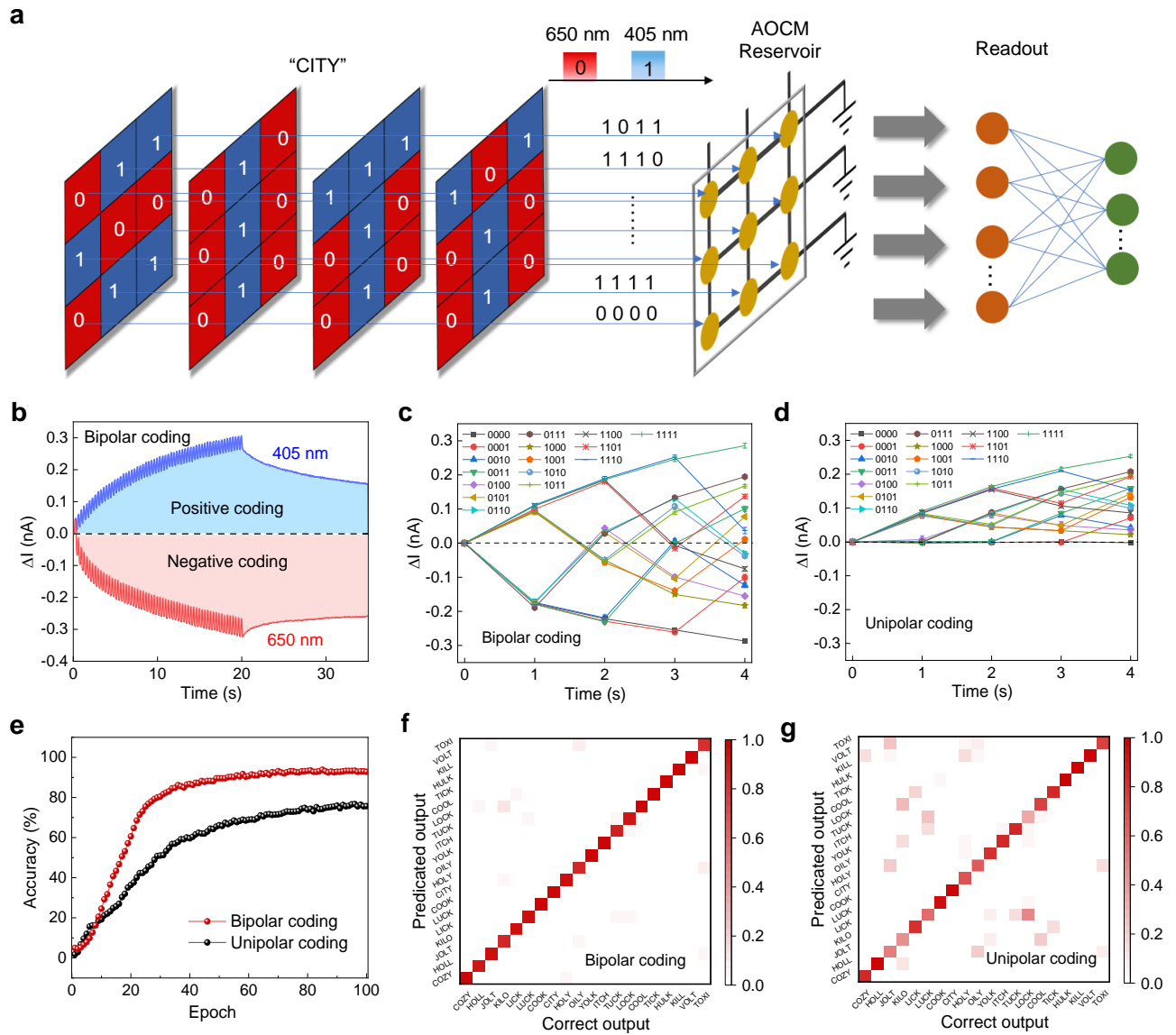

**Fig. 3 | Bipolar coding for word recognition based on AOCMs. a,** Schematic illustration of bipolar coding for word recognition using "CITY" as an example. Nine sets of 4-digit light pulse data streams are input into a reservoir layer composed of a 3×3 AOCM array. The resulting reservoir states are then fed into the readout layer of a single-layer ANN for recognition. **b,** Schematic of the bipolar coding region using 405 nm ($P$ = 0.8 mW/cm$^2$) and 650 nm light ($P$ = 10 mW/cm$^2$). **c,d,** Output reservoir states for 16 different input combinations under bipolar (**c**) and unipolar (**d**) coding modes. In bipolar coding, '1' represents a 405 nm light pulse ($P$ = 0.8 mW/cm$^2$, $W$ = 1 s), and '0' represents a 650 nm light pulse ($P$ = 10 mW/cm$^2$, $W$ = 1 s). No time interval exists between consecutive light pulses. In unipolar coding, '1' represents a 405 nm light pulse ($P$ = 0.8 mW/cm$^2$, $W$ = 1 s), and '0' represents no light pulse applied. **e,** Recognition accuracies under bipolar and unipolar coding modes. **f,g,** Confusion matrixes for bipolar (**f**) and unipolar (**g**) coding modes. Colour bar: occurrence probability of the predicted output.

Beyond language recognition, RC has found significant applications in temporal signal prediction – a fundamental approach for analyzing the dynamic behavior of complex systems, such as in climate modeling. The Lorenz system[50], a classic chaotic model describing atmospheric convection, serves as a standard benchmark in this field. Herein, we assess the time-series prediction performance of our bipolar coding-based AOC RC system using the Lorenz system, which is governed by the following ordinary differential equations: $dx/dt = \delta(y-x)$; $dy/dt = x(\rho-z)-y$; $dz/dt = xy-\beta z$, where x, y, z are state variables, t denotes time, and $\delta$, $\rho$, $\beta$ are system parameters[50]. From these equations, we generated a time-series dataset capturing the Lorenz system's dynamic trajectory. The dataset consists of 800 sequential data points scaled between –1 and 1 (sampled at a constant interval of 0.2). We divided it into 600 points for training and 200 for testing. To enhance reservoir dynamics, we employed a time-division multiplexing strategy, multiplying the raw signal by a binary mask (constructed from random combinations of 1 and –1) to generate virtual nodes in time domain[9]. The length (*L*) of the mask sequence has a critical impact on prediction accuracy: excessively long masks reduce inter-node correlation, whereas overly short ones limit state diversity. In this study, we experimented with masks of varying lengths, using a mask of length 5 as an illustrative example of the time-division multiplexing principle. As illustrated in **Fig. 4a**, each raw value (*e.g.*, 0.8) is converted into a higher-dimensional sequence of five values (0.8, –0.8, –0.8, 0.8, –0.8) by applying the mask sequence (1, –1, –1, 1, –1). These masked signals are then coded into light pulses: positive values correspond to 405 nm pulses, and negative values to 650 nm pulses. The signal amplitudes are linearly mapped to power densities, with 405 nm light spanning 0.2 to 1.0 mW/cm² in increments of 0.2 mW/cm² and 650 nm light ranging from 2 to 10 mW/cm² in increments of 2 mW/cm². Leveraging the dynamic and nonlinear bipolar photoresponse of our AOCMs, the generated virtual nodes become nonlinearly coupled with one another. **Figure 4b** displays the resulting ΔI under bipolar coding, showing reservoir states evolving between positive and negative values due to dual-wavelength coupling. For each time series signal, five reservoir states are selected as feature values for subsequent sequence prediction tasks, as indicated by the purple markers in **Fig. 4b**. For comparison, we also implemented unipolar coding, where negative values are represented by darkness (no illumination) and positive values by 405 nm light. **Figure 4c,d** summarize the prediction results under bipolar and unipolar coding, yielding average NRMSE values for the (x, y, z) dimensions of about 0.12 and 0.31, respectively – confirming the superiority of bipolar coding.

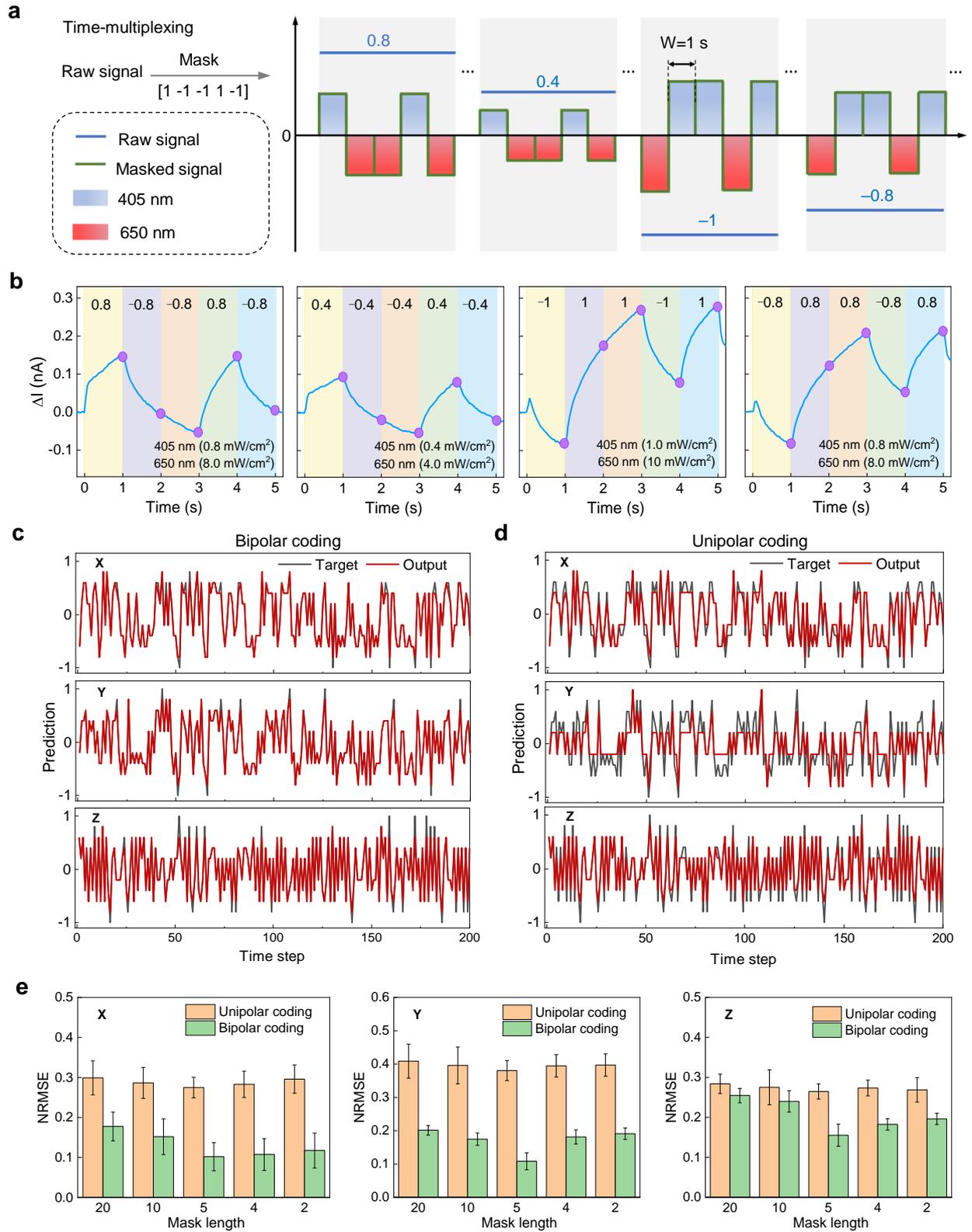

**Fig. 4 | Bipolar coding for time-series prediction. a,** Schematic of bipolar coding for time-series prediction. The figure shows the bipolar encoding process for a mask length of 5, where masked values are represented by 405 nm light pulses ($W = 1$ s) for positive values and 650 nm ($W = 1$ s) light pulses for negative values. There is no time interval between consecutive light pulses. The

amplitudes of the masked signals are linearly mapped to the values of *P*. **b,** Real-time ΔI for different masked signals. **c,d,** Prediction results for the x, y, and z dimensions under bipolar (c) and unipolar (d) coding. In unipolar coding, negative masked signal values are represented as darkness (no light), while positive values are represented by 405 nm light pulses. The parameters of the 405 nm light pulse are the same as those in (**b**). **e,** The relationship between NRMSE values and mask length *L*.

**Figure 4e** illustrates the impact of the mask length on NRMSE while keeping the total reservoir size ($L \times N$) fixed at 20, where *N* is the number of AOCM devices. We designed experimental groups with *L*= 20, 10, 5, 4, and 2, selecting *N*=1, 2, 4, 5, and 10 devices as parallel reservoirs, respectively. The results show that NRMSE increases when the mask length is either too long or too short, reaching its minimum value at a mask length of 5. Additionally, across all configurations, bipolar coding consistently outperforms unipolar coding.

In summary, across both spatial pattern recognition and chaotic time-series prediction, bipolar coding consistently and significantly outperforms the conventional unipolar approach. This performance gain is directly attributable to the richer, higher-dimensional reservoir dynamics enabled by the all-optical, bidirectional control of conductance states in our AOCM devices.

**Parallel Coding for Multi-Source Information Fusion**

Efficient multi-source information fusion technology enhances system security by integrating data from diverse sources, modalities, and dimensions[35]. A typical application is dual-factor authentication, which combines modalities like face and fingerprint recognition to enhance security in high-risk scenarios such as large financial transfers or payment password changes. However, conventional multi-source fusion approaches often involve complex ANNs that perform feature extraction and fusion in separate stages[34,35]. This separated architecture results in high computational overhead and hardware resource consumption, making it unsuitable for resource-constrained edge environments. To overcome these limitations, we propose and demonstrate a novel integrated fusion architecture based on our AOC RC system using parallel optical coding. This approach leverages the intrinsic physical coupling within a single AOCM device under simultaneous multi-wavelength illumination to to perform synchronous feature extraction and fusion, merging two processing stages into one unified physical operation.

We conducted a dual-authentication experiment combing face and fingerprint recognition as a representative case study. Ten pairs of face and fingerprint images were collected and converted into binary black-and-white images with 32×64 pixels to simplify computation (see **Supplementary Fig. 6** and **Methods**). For comparison, we executed the same authentication task using unipolar, bipolar, and parallel coding schemes based on the AOC RC system. **Figure 5a** illustrates the unipolar-input reservoir configuration, which utilizes the unidirectional photoresponse of AOCMs under 405 nm illumination. In this setup, both face and fingerprint data are coded using 405 nm light pulses, with feature extraction performed by two independent AOCM reservoirs before subsequent fusion and recognition. **Figure 5b** demonstrates the bipolar-input reservoir implementation, which leverages the bidirectional photoresponse of AOCMs through synchronized 405 nm and 650 nm illumination. Although this approach enables bipolar coding, it still necessitates two separate AOCM reservoirs and preserves a distinct separation between feature extraction and fusion stages. The proposed parallel-input reservoir architecture, depicted in **Fig. 5c**, achieves integrated feature extraction and fusion within a single reservoir through the synchronous processing of light signals.

In our implementation, we employed a 4-digit coding scheme where face information was coded as 405 nm light pulses and fingerprint information as 650 nm light pulses. These coded light signals were projected simultaneously onto the same AOCM device. Following the methodology established in our previous word recognition task using bipolar coding, we selected a power density combination of 0.8 mW/cm² (405 nm) and 8 mW/cm² (650 nm) to ensure a balanced distribution of reservoir states across both positive and negative coding regions, thereby maximizing the discriminative capability of the reservoir states.

To illustrate the coupling process of reservoir states under parallel coding mode, we present the real-time photocurrent evolution under different coding combinations (**Fig. 5d-f**). **Figure 5d** shows the temporal evolution of the photocurrent when both 405 nm and 650 nm light signals are coded as (1111). Herein, '1' corresponds to a light pulse. The photocurrent measured at 0.3 s after coding completion is taken as the resulting reservoir state, marked by the purple symbol in **Fig. 5d**. The coupled ΔI of approximately 0.02 nA differs significantly from the values of about 0.13 nA (under 405 nm illumination) and –0.16 nA (under 650 nm illumination), highlighting the coupling effect between current-enhancing and current-suppressing photoresponses under dual-wavelength illumination. When the 405 nm light is kept coded as (1111) and the 650 nm light coding is changed to (1010), the coupled ΔI increases to around 0.1 nA (**Fig. 5e**). Herein, '0' corresponds to darkness (without light). Conversely, when the 650 nm light remains coded as (1111) and the 405 nm coding is altered to (1010), the coupled ΔI decreases to about –0.05 nA (**Fig. 5f**). These results demonstrate that by regulating the proportion of 405 nm and 650 nm light pulses in the parallel coding scheme, the reservoir states can exhibit bipolar dynamic evolution.

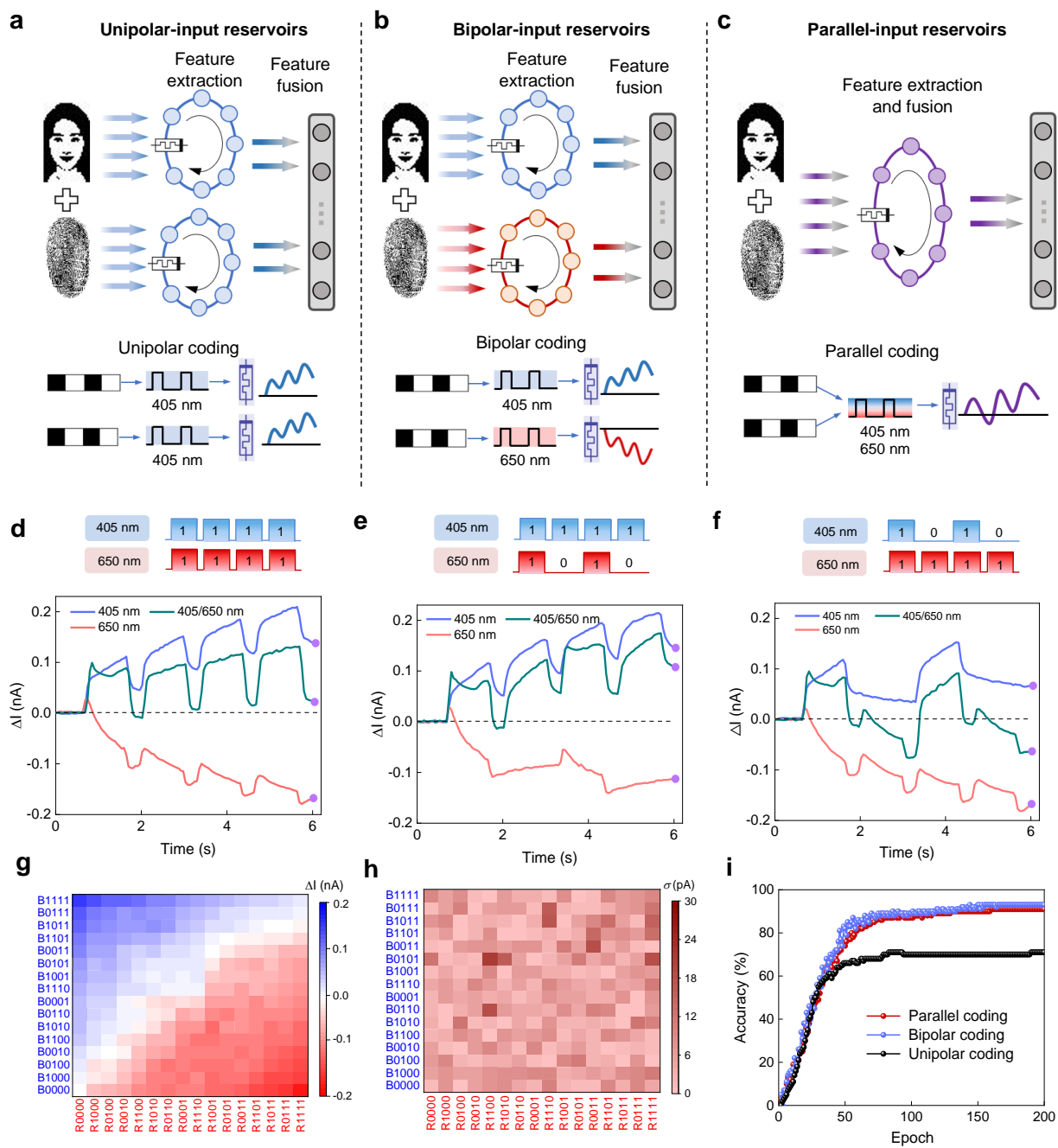

**Fig. 5 | Parallel coding for multi-source information fusion. a–c,** Schematics of multi-source information fusion based on unipolar (**a**), bipolar (**b**), and parallel (**d**) coding modes. In (**a,b**), feature extraction is carried out by reservoirs for input signals from different sources individually. The

extracted features are then fused and classified. In (**c**), feature extraction and multi-signal fusion are executed within a parallel-input reservoir. **d–f,** ΔI evolution process under the parallel coding mode with different combinations of 405 nm (*P* = 0.8 mW/cm$^2$, *W* = 1 s, *In* = 0.3 s) and 650 nm (*P* = 8 mW/cm$^2$, *W* = 1 s, *In* = 0.3 s) light pulses. For comparison, the figures also show the ΔI variations when exposed to individual 405 nm or 650 nm light pulses. **g**, Distribution maps of ΔI for all 4-digit parallel coding combinations. The testing method is consistent with (**d–f**). The statistical values of ΔI are the photocurrents at 0.3 s after parallel coding, as indicated by the purple markers in (**d–f**). **h**, Standard deviation maps of ΔI obtained after three repetitions of the parallel coding tests shown in (**g**). **i**, Recognition accuracies under unipolar, bipolar, and parallel coding.

We comprehensively evaluated the performance of parallel coding with 4-digit configurations by testing all 256 possible coding combinations. **Figure 5g** displays the ΔI distribution measured at 0.3 s after coding completion, clearly illustrating the bipolar dynamic evolution induced by dual optical coupling. **Figure 5h** presents the standard deviation ($\sigma$) for each coding combination, obtained from three repeated tests, to demonstrate the operational reliability of the parallel coding. The small $\sigma$ values indicate strong reliability of the parallel coding strategy. Based on the characterization results in **Fig. 5g,h**, datasets were generated representing reservoir-mapped outputs of face and fingerprint information for the dual-authentication task, comprising 1000 training samples and 9000 testing samples. **Figure 5i** compares the recognition performance of different coding schemes. Both bipolar and parallel coding achieve high accuracy, at 91% and 90.5%, respectively, significantly surpassing unipolar coding, which reaches only 70%. Notably, the parallel coding approach delivers accuracy comparable to that of bipolar coding while considerably reducing hardware demands and computational complexity. These technological advantages mainly stem from two aspects: first, bipolar coding requires two separate reservoirs, while parallel coding operates with only one; second, bipolar coding produces 1024 feature values per face-fingerprint pair, compared to only 512 in the parallel scheme (see **Methods** for details).

**Conclusion**

We have demonstrated an AOC RC system based on a ZnO-AOCM array that directly addresses the critical limitations of dynamic range and nonlinearity in conventional optoelectronic RC systems. The platform leverages the unique wavelength-dependent photoresponse of fabricated ZnO devices, where selected illumination with blue (405 nm) and red (650 nm) light induces positive and negative persistent photoconductivity, respectively. This bidirectional photomodulation originates from the dynamic evolution of oxygen vacancies in ZnO. By tuning the power density and irradiation mode of dual-wavelength light pulses, we achieve controllable modulation of photocurrent relaxation and nonlinear dynamics. Building on these properties, we introduce bipolar and parallel coding strategies that substantially expand reservoir state diversity and enhance nonlinear mapping capacity. Compared to traditional unipolar coding, the bipolar coding shows significantly enhanced accuracy in word recognition and time-series prediction tasks. The parallel coding facilitates efficient feature extraction and fusion of multi-source signals within a single reservoir, retaining high computational accuracy while considerably lowering hardware consumption. Overall, the proposed AOCM-based architecture provides a promising platform for high-performance RC and presents a feasible pathway toward the compact, hardware-level realization of edge intelligent systems.

## Methods

### Array fabrication and characterization

The 16×16 device array with a crossbar structure was fabricated using *in situ* metal shadow masks. First, Ti (~15 nm) and Pt (~10 nm) layers were sequentially deposited on a quartz substrate by RF and DC magnetron sputtering, respectively, using a striped shadow mask with a line width of 50 μm. This process produced 16 parallel Pt wires serving as the bottom electrodes of the device. Next, ZnO thin films (~80 nm) were deposited on the Pt/Ti/quartz substrates via RF magnetron sputtering at room temperature (RT) in a pure argon atmosphere, employing a 99.99% purity ZnO ceramic target. The sputtering power of the ZnO films was 60 W, with a sputtering pressure of 0.5 Pa. During deposition, a shadow mask featuring 16×16 square openings with a side length of 100 μm was used, yielding 16×16 distinct, approximately square ZnO thin film regions. The ZnO film was then annealed in air at 600 °C for 1 hour. Finally, the top Pt electrodes (~10 nm) were deposited on the ZnO/Pt/Ti/quartz structure by RF magnetron sputtering using the same striped mask as for the bottom electrodes, rotated by 90°. The thickness of each layer was measured using a variable angle spectroscopic ellipsometer (M-2000 DI, J. A. Woollam Co., Inc.).

Electrical and optoelectronic characterizations were performed at RT in air using a semiconductor parameter analyzer (Primarius FS-Pro) equipped with a monochromatic light source (Omni-λ 3007), a 405 nm laser, and a 650 nm laser. The monochromatic light source provided 350 nm illumination. An arbitrary waveform generator (Liquid Instruments, Moku:Go) controlled the alternating and synchronized emission of 405 nm and 650 nm light pulses. During all electrical and optoelectronic measurements, the top electrode was biased, and the bottom electrode was grounded. Illumination was directed onto the memristive device through the top electrode.

**Simulation**

In the word recognition task (**Fig. 3**), nine reservoir states were collected for each word and then input into the subsequent readout layer. The readout layer was a 9×20 network realized in Python, with 20 outputs corresponding to 20 different words. Based on the characterization results shown in **Fig. 3c** and **Supplementary Fig. 4**, we generated a dataset representing reservoir-mapped outputs of English words, comprising 1000 training samples and 1000 testing samples.

For the time-series prediction task in **Fig. 4**, the mask values were randomly generated using Python. The generated mask sequences of length 2 were: (1, –1), (1, –1), (–1, 1), (1, –1), (1, –1), (–1, 1), (–1, 1), (1, –1), (–1, 1), (1, –1). The generated sequences of length 4 were: (–1, 1, –1, –1), (–1, 1, –1, 1), (1, –1, 1, 1), (1, –1, 1, –1), (–1, 1, –1, –1), (1, 1, 1, –1). The sequences of length 5 were: (1, –1, –1, 1, –1), (1, 1, –1, –1, –1), (–1, –1, 1, –1, 1), (–1, 1, 1, –1, –1). The sequences of length 10 were: (–1, 1, 1, –1, –1, 1, –1, 1, –1, 1), (–1, –1, 1, 1, –1, –1, –1, 1, 1, –1). The sequences of length 20 were: (1, –1, –1, 1, 1, –1, 1, 1, –1, –1, 1, 1, 1, –1, 1, 1, 1, –1, –1, 1). In the Lorentz system, the parameters $\delta$, $\rho$, and $\beta$ are typically set to 10, 28, and 8/3, respectively[50].

For the multi-source information fusion task illustrated in **Fig. 5**, we collected ten pairs of face and fingerprint images, which were subsequently converted into binary black-and-white images using Python. The task employed a 4-digit coding scheme where each coding operation yielded one reservoir state. Following feature extraction, each pair of face and fingerprint images generated 1024 distinct feature values, which were then fed into the readout layer. Accordingly, in **Fig. 5a,b**, the readout layer was implemented as a 1024×100 network, with the 100 output nodes corresponding to the 100 possible combinations of face and fingerprint identities. In the parallel coding configuration shown in **Fig. 5c**, the integrated processing of each face-fingerprint pair produced 512 feature values after feature extraction. Consequently, the readout layer in **Fig. 5c** was structured as a 512×100 network.

In the tasks presented in **Figs. 3-5**, a sigmoid function was used as the activation function of the readout network to calculate the probabilities of different outputs. The cost function was minimized through the gradient descent method.

**Data availability**

All other data are available from the corresponding authors upon reasonable request.

**Code availability**

The codes used for simulation and data plotting are available from the corresponding authors upon reasonable request.

**Acknowledgements**

This research was partially supported by the National Natural Science Foundation of China (Nos. 62304228 and U25A20498), the Zhejiang Provincial Natural Science Foundation of China (No. LD25F040006), the Ningbo Municipal Science and Technology Innovation Yongjiang 2035 Key

R&D Plan (No. 2025Z086), and the Ningbo Global Innovation Center, Zhejiang University.

**Competing interests**

The authors declare no competing interests.

**Additional information**

Supplementary information: the online version contains supplementary material available at https://doi.org/

# Supporting Information

**All-Optically Controlled Memristive Reservoir Computing Capable of Bipolar and Parallel Coding**

Lingxiang Hu[1 #], Dian Jiao[1 #], Peihong Cheng[1,2], Jingrui Wang[1,2], Li Zhang[3], Athanasios V. Vasilakos[4], Yang Chai[5], Zhizhen Ye[6,7] & Fei Zhuge[1,6,8,9] *

[1]Ningbo Institute of Materials Technology and Engineering, Chinese Academy of Sciences, Ningbo, China

[2]School of Electronic and Information Engineering, Ningbo University of Technology, Ningbo, China

[3]Healthcare Engineering Centre, School of Engineering, Temasek Polytechnic, Singapore

[4]Center for AI Research, University of Agder, Grimstad, Norway

[5]Department of Applied Physics, The Hong Kong Polytechnic University, Hong Kong, China

[6]Institute of Wenzhou, Zhejiang University, Wenzhou, China

[7]State Key Laboratory of Silicon and Advanced Semiconductor Materials, School of Materials Science and Engineering, Zhejiang University, Hangzhou, China

[8]Center for Excellence in Brain Science and Intelligence Technology, Chinese Academy of Sciences Shanghai, China

[9]College of Materials Science and Opto-Electronic Technology, University of Chinese Academy of Sciences, Beijing, China

[#]L.H. and D.J. contributed equally to this paper.

*E-mail: zhugefei@nimte.ac.cn

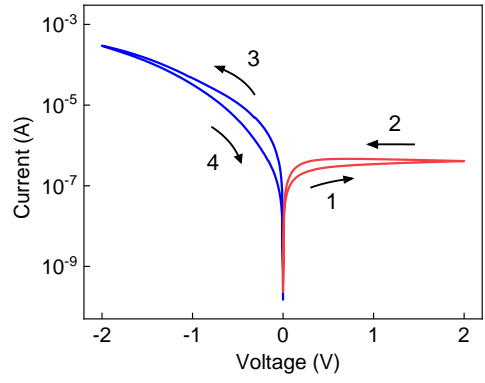

**Supplementary Figure 1.** Current–voltage characteristics of the AOCM in the dark.

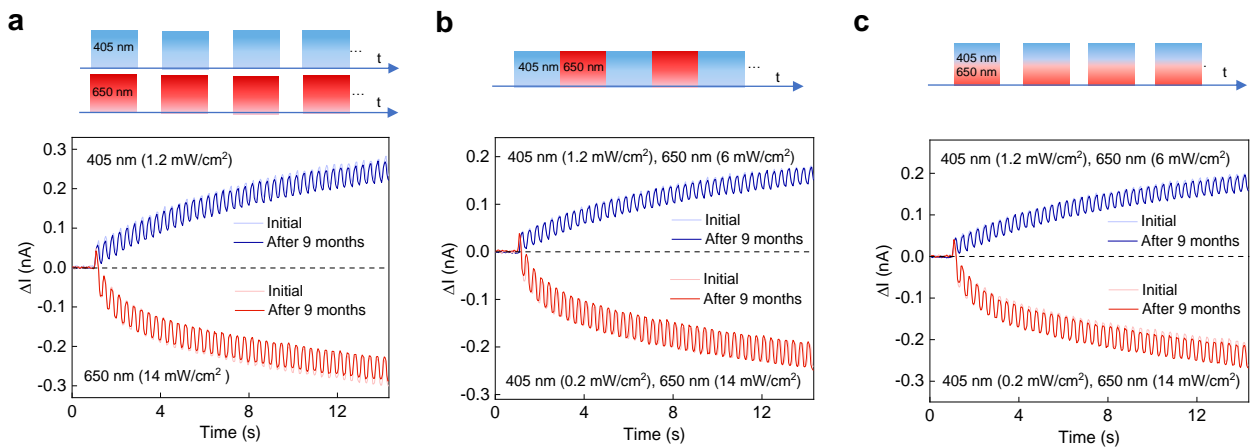

**Supplementary Figure 2.** Stability of the bipolar photoresponse in the AOCM device over time. The bipolar photoresponse remains stable after 9 months under illumination schemes of (a) individual, (b) alternating, and (c) synchronous 405 nm and 650 nm light pulses. In (a), the parameters for the 405 nm and 650 nm light pulses are: $W$ = 0.15 s, $In$ = 0.15 s, $N$ = 40). The illumination schemes in (b) and (c) correspond to those shown in Fig. 2f,i, respectively. Devices were vacuum-sealed and stored at room temperature.

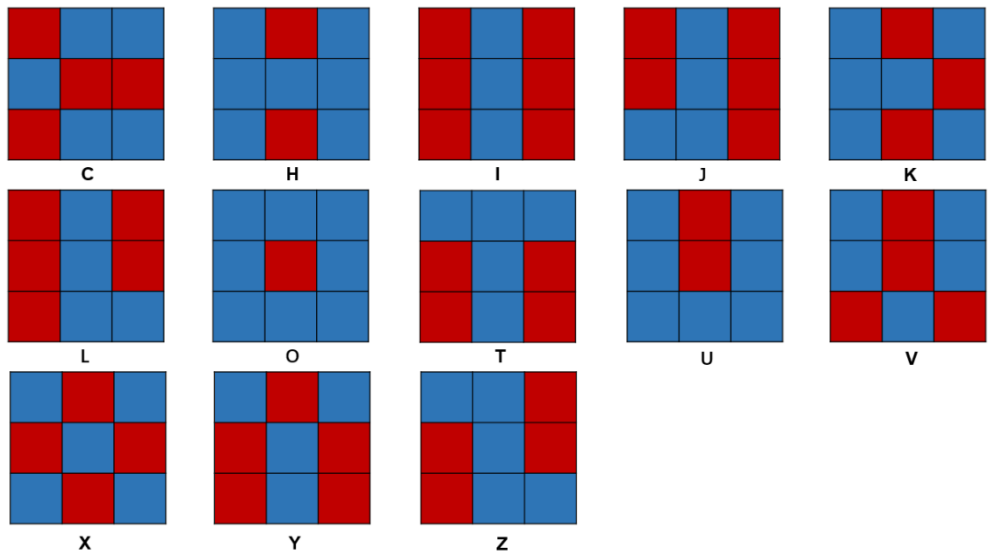

**Supplementary Figure 3.** Schematics of 13 letters with a 3×3 pixel layout: 'C', 'H', 'I', 'J', 'K', 'L', 'O', 'T', 'U', 'V', 'X', 'Y', 'Z'. These letters can form 20 different four-letter English words: 'COZY', 'HOLL', 'JOLT', 'KILO', 'LICK', 'LUCK', 'COOK', 'CITY', 'HOLY', 'OILY', 'YOLY', 'ITCH', 'TUCK', 'LOCK', 'COOL', 'TICK', 'HULK', 'KILL', 'VOLT', 'TOXI'.

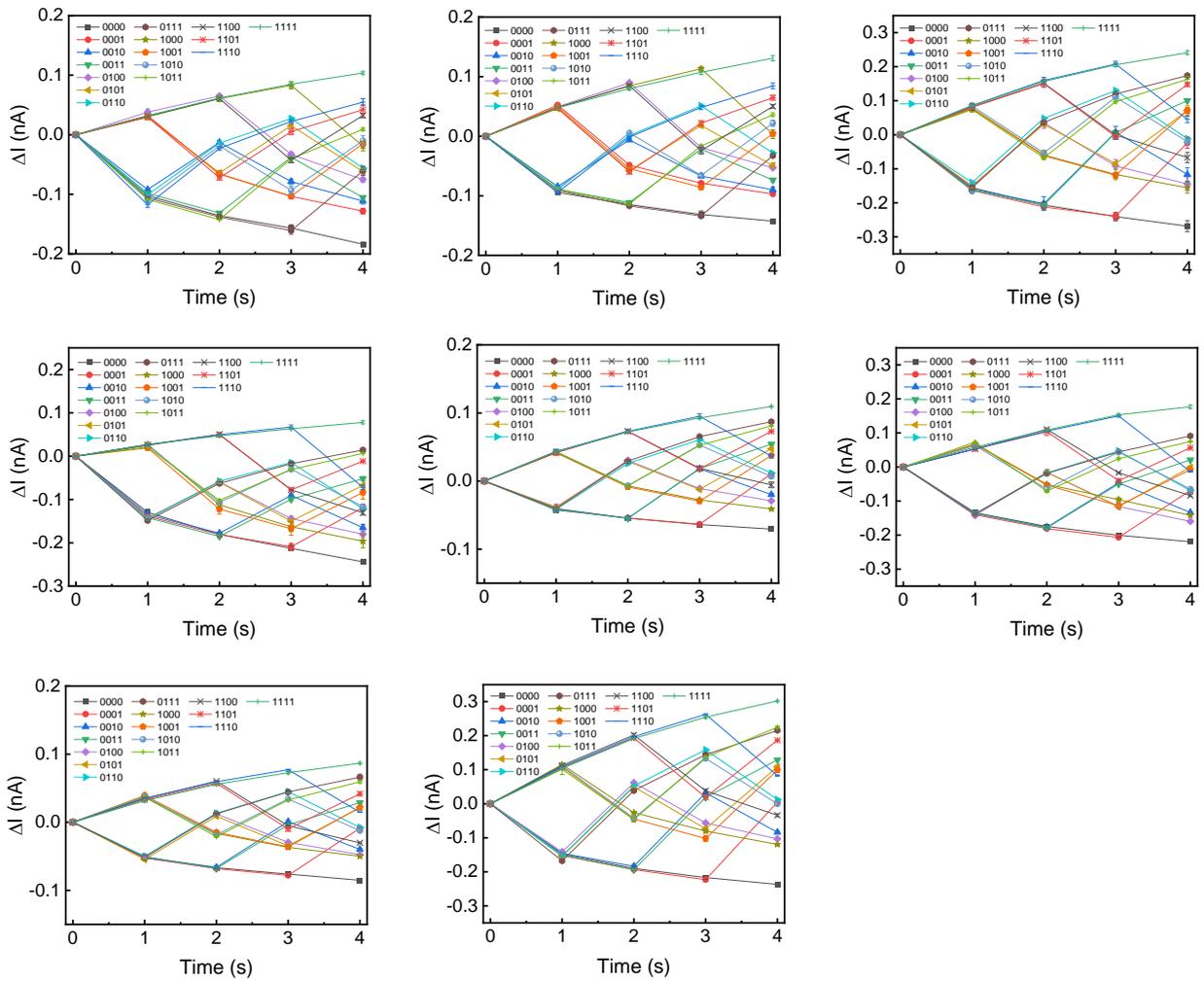

**Supplementary Figure 4.** The bipolar coding results of the remaining eight AOCM devices for the word recognition task in Fig. 3. In bipolar coding, '1' represents applying a 405 nm light pulse ($P$ = 0.8 mW cm$^{-2}$, $W$ = 1 s) , while '0' represents applying a 650 nm light pulse ($P$ = 10 mW cm$^{-2}$, $W$ = 1 s).

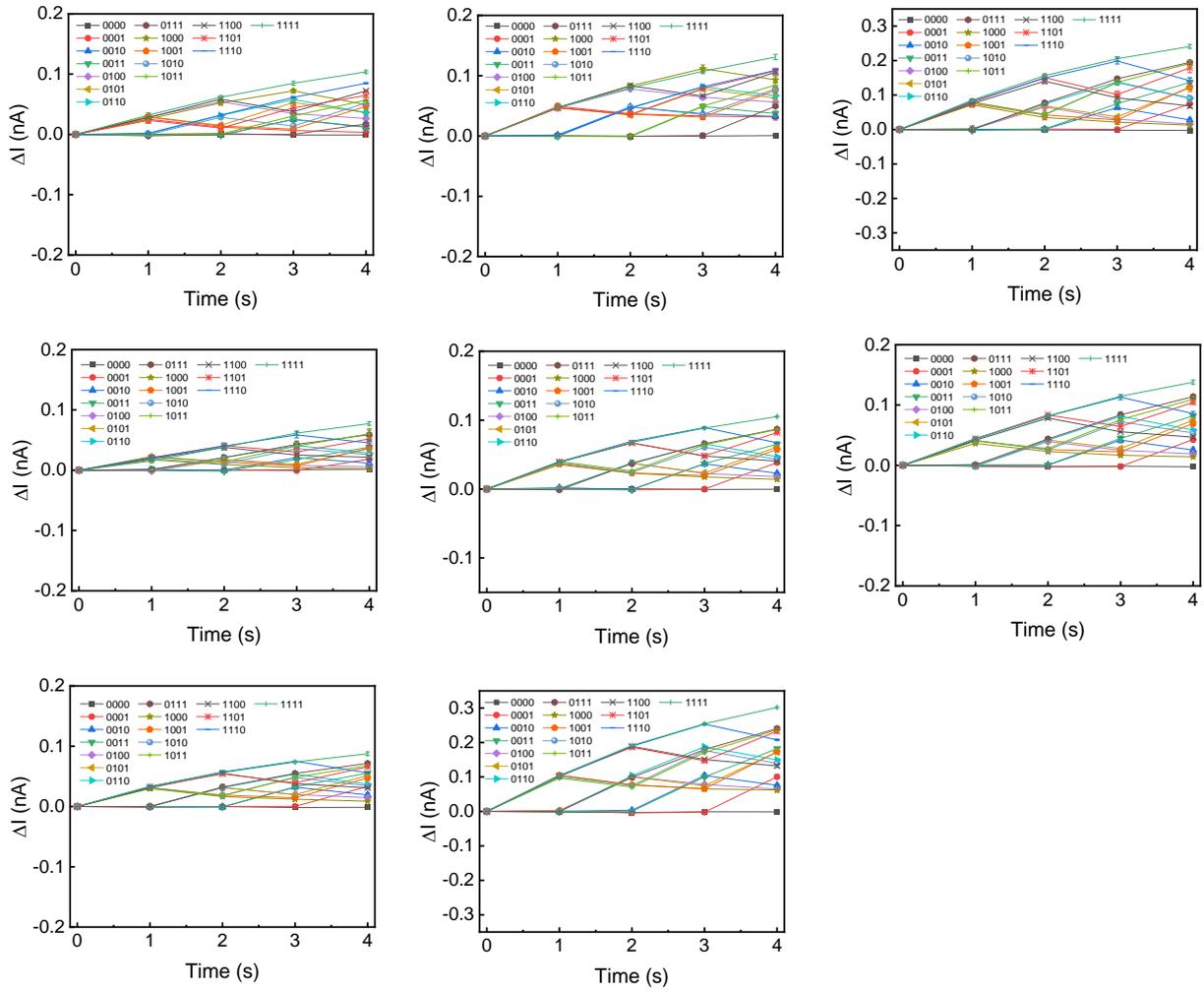

**Supplementary Figure 5.** The unipolar coding results of the remaining eight AOCM devices for the word recognition task in Fig. 3. In unipolar coding, '1' represents applying a 405 nm light pulse ($P$ = 0.8 mW cm$^{-2}$, $W$ = 1 s), while '0' indicates that no light pulse is applied.

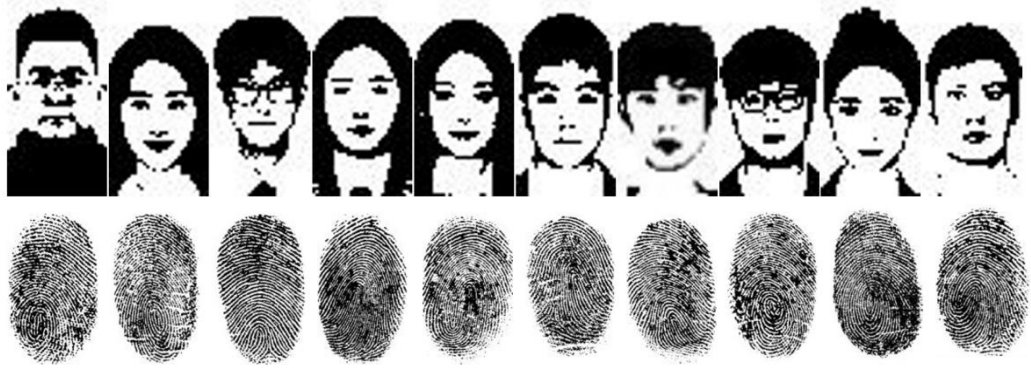

**Supplementary Figure 6.** Ten pairs of face and fingerprint images were collected for the multi-source information fusion task presented in Fig. 5.